\documentclass[10pt,conference]{IEEEtran}

\IEEEoverridecommandlockouts
\usepackage{cite}
\usepackage{amsmath,amssymb,amsfonts}
\usepackage{algorithmic}
\usepackage{graphicx}
\usepackage{textcomp}
\usepackage{xcolor}

\newtheorem{definition}{Definition}
\newtheorem{assumption}{Assumption}

\newtheorem{theorem}{Theorem}

\usepackage[colorlinks=true,linkcolor=blue]{hyperref}

\usepackage{xspace}
\usepackage{wrapfig}
\usepackage{subfig}
\usepackage{bm}
\usepackage{listings}
\lstdefinelanguage{LF}{
  keywords={deadline, after, state, logical, physical, startup, shutdown, reaction, preamble, target, reactor, trigger, input, output, constructor, new, action, clock, actor, handler, time, main, federated, timer, sec, secs, msec, msecs, usec, usecs, ms},
  emph={L,name, type, init, effect, instance}, emphstyle=\itshape,
  keywordstyle=\color{black}\bfseries,
  ndkeywords={class, export, boolean, throw, implements, import, this},
  ndkeywordstyle=\color{darkgray}\bfseries,
  identifierstyle=\color{black},
  sensitive=false,
  comment=[l]{//},
  morecomment=[s]{/*}{*/},
  commentstyle=\color{purple}\ttfamily,
  stringstyle=\color{black}\ttfamily,
  morestring=[b]',
  morestring=[b]"
}
\newboolean{showcomments}
\setboolean{showcomments}{false} %set false for final submission
\newcommand{\fixme}[1]{
	\ifthenelse{\boolean{showcomments}}
	{{\color{red}\textbf{FIXME:} #1}}
	{}
}
\newcommand{\physicalConn}{\texttt{{\hbox{$\small\mathtt{\sim}$}}>}\xspace}

\newcommand{\lfshort}[0]{\textsc{LF}\xspace}

\newcommand{\lf}[0]{\textsc{Lingua Franca}\xspace}

\newcommand{\lfdef}[0]{\lf (\lfshort)\xspace}

\begin{document}

\title{Consistency vs.~Availability in \\
Distributed Real-Time Systems*
\thanks{{\footnotesize \textsuperscript{*} This work was supported by the iCyPhy Research Center at UC Berkeley, supported by Denso, Siemens, and Toyota.}}
}

%\author{\IEEEauthorblockN{Anonymous}
%\IEEEauthorblockA{\textit{Anonymous} \\
%Anonymous \\
%Anonymous}
%}

\author{\IEEEauthorblockN{Edward A. Lee}
\IEEEauthorblockA{\textit{EECS, UC Berkeley} \\
Berkeley, USA}
\and
\IEEEauthorblockN{Ravi Akella}
\IEEEauthorblockA{\textit{Denso NA} \\
San Jose, USA}
\and
\IEEEauthorblockN{Soroush Bateni}
\IEEEauthorblockA{\textit{EECS, UC Berkeley} \\
Berkeley, USA} \\
\and
\IEEEauthorblockN{Shaokai Lin}
\IEEEauthorblockA{\textit{EECS, UC Berkeley} \\
Berkeley, USA}
\and
\IEEEauthorblockN{Marten Lohstroh}
\IEEEauthorblockA{\textit{EECS, UC Berkeley} \\
Berkeley, USA}
\and
\IEEEauthorblockN{Christian Menard}
\IEEEauthorblockA{\textit{TU Dresden} \\
Dresden, Germany}
}

\maketitle

\begin{abstract}
In distributed applications, Brewer's CAP theorem tells us that when networks
become partitioned (P), one must give up either consistency (C) or availability (A).
Consistency is agreement on the values of shared
variables; availability is the ability to respond to reads and
writes accessing those shared variables. Availability is a real-time property
whereas consistency is a logical property.
We have extended the CAP theorem to relate quantitative measures of these
two properties to quantitative measures of communication and computation latency (L),
obtaining a relation called the CAL theorem that
is linear in a max-plus algebra. This paper shows how to use the CAL theorem
in various ways to help design real-time systems.
We develop a methodology for systematically trading off
availability and consistency in application-specific ways and to guide
the system designer when putting functionality in end devices, in edge computers,
or in the cloud.
We build on the \lf coordination language to provide system
designers with concrete analysis and design tools to make the required tradeoffs
in deployable software.
\end{abstract}


\section{Introduction}

Brewer's well-known CAP Theorem states that in the presence of network partitioning (P),
a distributed system must sacrifice at least one of availability (A) or consistency (C)~\cite{Brewer:00:CAP,Brewer:12:CAP}.
Consistency is where distributed components agree on the value of shared state, and
availability is the ability to respond to user requests using and/or modifying that shared state.
Gilbert and Lynch \cite{GilbertLynch:02:CAP} prove two variants of this theorem,
one strong result for asynchronous networks~\cite[chapter 8]{Lynch:96:IOAutomata}
and one weaker result for partially synchronous networks.
The CAP theorem has helped the research and development of distributed
database systems by clarifying a fundamental limit
and suggesting application-dependent tradeoffs.
For some database applications, availability is more
important than consistency, while for others it is the other
way around.
The purpose of this paper is to apply
and adapt the CAP theorem to distributed real-time systems
to derive similar benefits.

Real-time systems are cyber-physical systems (CPSs) with timing constraints.
In distributed real-time systems, the state information shared between software components
is often information about the physical world in which the software is operating.
In autonomous vehicles, for example, the state of an intersection is shared among all the vehicles
contending for access to that intersection.
Even within a single vehicle, where software components may be distributed across an onboard network,
many of these software components will share state information about the vehicle and its environment.
We therefore generalize the notion of consistency to include physical state rather than just variables in software.

In real-time systems, the time it takes for a software subsystem to respond through an actuator to stimulus from a sensor
is a critical property of the system. We therefore generalize the notion of availability to include this time,
not just the system's response time to human users.
A software subsystem where sensor-to-actuator response time is large is less available than one for which it is small.

Brewer's CAP theorem, then, immediately applies in an obvious way.
When the network becomes partitioned, one of availability or consistency must be sacrificed.
However, network partitioning is just the limiting case of network latency, as pointed out by Brewer~\cite{Brewer:12:CAP} and
Abadi~\cite{Abadi:12:CAP}.
Moreover, network latency is not the only latency that can force this compromise.
Long execution times can be just as damaging as long network latencies, and so can large clock synchronization
discrepancies, as we will show.
In our formulation, network latencies, execution times, and clock synchronization errors get lumped together
into a single measurable quantity that we call, simply, ``latency'' (or ``apparent latency'' to emphasize that the quantity we work with is measurable).

We have recently discovered that consistency, availability, and latency can all be quantified,
and that they have a simple algebraic relationship between them.
We call this relationship the CAL theorem, replacing ``Partitioning'' with ``Latency.''
The relation is a linear system of equations in a max-plus algebra, where the structure of the equations reflects the communication topology of an application.
%To make this paper self contained, we review the CAL theorem here.

The only prior attempt we are aware of to quantify the CAP theorem was done by
Yu and Vahdat, who quantified availability and consistency and show a tradeoff between them~\cite{YuVahdat:06:CAP}.
Their quantifications, however, are in terms of fractions of satisfied accesses (availability) and fractions of out-of-order writes (inconsistency),
and they show that finding the availability as a function of consistency is NP-hard. In contrast, the CAL theorem defines these quantities as time intervals
and gives a strikingly simpler relationship, one that is linear in a max-plus algebra. 

How to trade off consistency, availability, and latency against one another is application specific.
Consider for example a four-way intersection, access to which is regulated by a distributed algorithm
running in software on autonomous vehicles that contend for the intersection.
For this application, and specifically for the state of the intersection, consistency is paramount.
All vehicles \emph{must} agree on the state of the intersection (strong consistency) before any one
vehicle can enter the intersection. Hence, for this application, when latencies get large for any reason,
we choose to sacrifice availability (vehicles do not enter the intersection) rather than consistency (vehicles crash).

Consider, however, a complementary application. Suppose a vehicle has a computer-vision-based automatic braking system
as part of an ADAS (advanced driver assistance system) as well as an ordinary brake pedal.
Suppose the vision-based system has significant latency (it may even be computed in the cloud).
Should the system delay responses to pushes of the brake pedal until the vision system has reported the state
of the world at the time of the brake pedal push?
The answer is most certainly ``no.''
The system should respond immediately to brake pedal pushes, thereby maintaining high availability,
even at the cost of consistency.

The CAL theorem also easily accommodates tiered, heterogeneous networks, where end devices
may connect to edge computers over wired or wireless links, and
edge computers may connect to cloud-based services that enable wide area
aggregation and scalability, for example for machine learning.
The various networks involved may have widely varying characteristics, yielding enormously different
latencies and latency variability.
A time-sensitive network (TSN)~\cite{LoBelloEtAl:19:TSN} on a factory floor, for example,
may yield reliable latencies on the scale of microseconds between edge computers,
whereas wide-area networks (WAN) may yield highly variable latencies that can extend up
to tens of seconds~\cite{ZhangEtAl:18:AWStream}.
Moreover, any of these networks can fail, yielding unbounded latencies, and systems need
to be designed to handle such failures gracefully.

The CAL theorem will allow us to model a heterogeneous network topology interconnecting
a wide variety of nodes. 
In particular, the matrix form of the equations enables compact modeling of heterogeneous networks,
where the latencies between pairs of nodes can vary considerably.

We will use the \lfdef coordination language
\cite{LohstrohEtAl:21:Towards} to specify programs that explicitly define availability and
consistency requirements for a distributed CPS application.
We can then use the CAL theorem to derive the network latency bounds that make meeting
the requirements possible. This can be used to guide decisions about which services must be
placed in the end devices, which can be placed on an edge computer, and which can be put in the cloud.
Moreover, we will show how, once such a system is deployed, violations of the network latency
requirements, which will make it impossible to meet the consistency and availability requirements,
can be detected. System designers can build in to the application fault handlers that handle such
failures.

System designers can use the CAL theorem in at least two complementary ways.
They can derive networking requirements from availability and/or consistency requirements,
or they can derive availability and/or consistency properties from assumptions about the network behavior.

The CAP theorem itself is rather obvious and very much part of the folklore in distributed computing.
By quantifying it and relating it to individual point-to-point latencies, the CAL theorem elevates
the phenomenon from folklore to an engineering principle, enabling rigorous design with clearly stated assumptions.
Moreover, by quantifying consistency and availability, the CAL theorem makes the concept applicable to real-time systems.
In this paper, we show how to carry out such rigorous design using \lfshort, %\cite{LohstrohEtAl:21:Towards} 
which supports explicit representations of availability and consistency requirements.
Moreover, we demonstrate how to detect situations where the networking requirements that are implied by the
availability and consistency requirements cannot be met, for example when the network fails or has excessive latency.
We describe how \lfshort can provide exception handlers that enable the designer to explicitly choose
how to handle such fault conditions, for example, by reducing accuracy~\cite{ZhangEtAl:18:AWStream}
or by switching to failsafe modes of operation.

The contributions of this paper are as follows:
\begin{itemize}
\item We show how availability, a real-time property of a system,
and consistency, a logical property, relate numerically to
clock synchronization and latencies introduced by networks and computation.
\item We derive how the deadlines commonly used to specify real-time requirements
in real-time systems
are availability requirements and therefore are subject to this relation.
Specifically, as latencies increase, it becomes impossible to meet deadlines without
sacrificing consistency.
\item We propose a methodology that allows a system designer to explicitly define availability and consistency
requirements using the \lf coordination language.
\item We illustrate how a system designer can choose how to handle runtime violations
of these requirements by explicitly choosing whether to further relax consistency or
availability  and how to provide fault handlers to be invoked when violations are detected.
\item We give practical real-time systems examples that show that the choice of whether to
sacrifice availability or consistency when faults occur is application dependent.
\end{itemize}

The paper is organized as follows.
Section~\ref{sec:time} explains the underlying model of time.
Section~\ref{sec:cal} formally defines the terms and derives the CAL theorem.
Section~\ref{sec:lf} introduces the \lf coordination language, shows how it can explicitly specify availability
and consistency requirement, and shows how to use the CAL theorem to analyze a program.
Section~\ref{sec:tradeoffs} gives two practical distributed real-time system examples and shows
how one needs to prioritize availability while the other needs to prioritize consistency in the presence of faults.
Section~\ref{sec:conclusion} draws some conclusions.

%%%%%%%%%%%%%%%%%%%%%%%%%%%%%%%%%%%%%%%%%%%%%%%%%%%%%%%%
\section{Logical and Physical Time}\label{sec:time}

Central to our ability to quantify both consistency and availability is the use of
two distinct notions of time, \textbf{logical} and \textbf{physical}.
A physical time $T \in \mathbb{T}$ is an imperfect measurement of time 
taken from some clock somewhere the system.
The set $\mathbb{T}$ contains all the possible times that a physical clock can report.
We assume that $\mathbb{T}$ is totally ordered and includes two special members:
$\infty, -\infty \in \mathbb{T}$, larger and smaller than any time any clock can report. We will occasionally make a distinction between the set $\mathbb{I}$
of time \textit{intervals} (differences between two times) and time \textit{values} $T \in \mathbb{T}$.
It is often convenient to have the set $\mathbb{T}$ represent a common definition of physical time, such as Coordinated Universal Time (UTC)
so that times correlate with physical reality.

In \lfshort, which we will use to specify our real-time systems, 
$\mathbb{T}$ and $\mathbb{I}$ are both
the set of 64-bit integers (for all targets built to date).
Following the POSIX standard, $T \in \mathbb{T}$ is the number of nanoseconds that have elapsed since 
midnight, January 1, 1970, Greenwich mean time.
The largest and smallest 64-bit integers represent $\infty$ and $-\infty$, respectively.
As a practical matter, these numbers will overflow in systems running near the year 2270.

For \textbf{logical time}, we use an element that we call a \textbf{tag} $g$ of a totally-ordered set $\mathbb{G}$.
Each event in a distributed system is associated with a tag $g \in \mathbb{G}$.
From the perspective of any component of a distributed system, the order in which events occur is defined by the order of their tags.
If two distinct events have the same tag, we say that they are \textbf{logically simultaneous}.
We assume the tag set $\mathbb{G}$ also has largest and smallest elements.
Moreover, we assume a metric that measures the distance between two tags.
This metric is what we will use to quantify consistency.

In \lfshort,
$\mathbb{G} = \mathbb{T} \times \mathbb{U}$, where $\mathbb{U}$ is the set of 32-bit unsigned integers representing the microstep of a superdense time system~\cite{Maler:92:Hybrid,Cataldo:06:Tetric,CremonaEtAl:17:Hybrid}.
Following the tagged-signal model~\cite{LeeSan:98:TaggedSignal},
we use the term \textbf{tag} rather than timestamp to allow for such a richer model of logical time.
For the purposes of this paper, however, the microsteps will not matter, and hence you can think of a tag as a timestamp and ignore the microstep.
In particular, the metric we will use to measure the distance between tags ignores the microstep.

We will consistently denote tags with a lower case $g \in \mathbb{G}$ or a lower-case
tuple $(t, m) \in \mathbb{G}$ and measurements of physical time $T \in \mathbb{T}$ with upper case.

To combine tags with physical times, we assume a monotonically nondecreasing function $\mathcal{T}\colon \mathbb{G} \to \mathbb{T}$
that gives a physical time interpretation to any tag.
For any tag $g$, we call $\mathcal{T}(g)$ its \textbf{timestamp}.
In \lfshort, for any tag $g = (t,m) \in \mathbb{G}$, $\mathcal{T}(g) = t$.
Hence, to get a timestamp from a tag, you just have to ignore the microstep.
The set $\mathbb{G}$ also includes largest and smallest elements such that
$\mathcal{T}(\infty_{\mathbb{G}}) = \infty_{\mathbb{T}}$
and
$\mathcal{T}(-\infty_{\mathbb{G}}) = -\infty_{\mathbb{T}}$,
where the subscripts disambiguate the infinities.

An external input, such as a user input or query, will be assigned a tag $g$ such that
$\mathcal{T}(g) = T$,
where $T$ is a measurement of physical time taken from the local clock where the input
first enters the software system.
In \lfshort, this tag is normally given microstep 0, $g = (T,0)$.

For any tag $g$, the time $\mathcal{T}(g)$ is a \textbf{logical time}.
It may be \emph{derived from} a physical time, as it is for an externally triggered event, but
once the tagged event enters the system, its relationship to physical time becomes incidental.
The only requirement is that software components process events in tag order, irrespective of
physical time.
We will use the tags $g$ to specify consistency requirements.
In \lf, however, to get real-time behavior, it is possible to associate \textbf{deadline} with the processing of an event.
A deadline $d \in \mathbb{I}$ is a declaration that the event must be processed before physical time $T$
exceeds the logical time $\mathcal{T}(g) + d$.
We will use these deadlines to specify availability requirements.

%%%%%%%%%%%%%%%%%%%%%%%%%%%%%%%%%%%
\section{The CAL Theorem} \label{sec:cal}

Following Schwartz and Mattern \cite{SchwarzMattern:94:CausalConsistency},
assume we are given a \textbf{trace} of an execution of a distributed system consisting of $N$ sequential \textbf{processes},
where each process is an unbounded sequence of (tagged) \textbf{events}.
Although the theory is developed for traces, the CAL theorem can be used for \emph{programs}, not just traces because
a program is formally a family of traces.
The $k$-th event of a process
is associated with a tag $g_k$ and a physical time $T_k$.
The physical time $T_k$ is the reading on a local clock at the time where the event starts being processed.
The events in a process are required to have nondecreasing tags and increasing physical times.
That is, if $g_k$ is the tag and $T_k$ is the physical time of the $k$-th event, then
$g_k \le g_{k+1}$ and $T_k < T_{k+1}$.

Within each process,
a \textbf{read event} with tag $g$ yields the value of a shared variable $x$.
The shared variable $x$ is stored as a local copy,
which has previously acquired a value via
a \textbf{write event} or 
an \textbf{accept event} in the same process.
An accept event receives an updated value of the variable from the network.
A read event with tag $g$ will yield the value assigned by the write or accept event with the largest tag $g'$ where $g' \le g$.
If $g' = g$, we require that $T' < T$, where $T'$ is the physical time of the write or accept event
and $T$ is the physical time of the read event.
This requirement ensures that a read event reads a value that was written at an earlier physical time.

A \textbf{send event} is where a process launches into the network an update to a shared variable $x$.
(Here, the ``network'' is whatever medium is being used for communication between processes.)
Like a read event, the send event has a tag greater than or equal to that of the write or accept event that it is reporting
and a physical time greater than that of the write or accept event.
An accept event that receives the update sent by the send event has a tag greater than or equal to that of the send event. 
The physical time of the accept event relative to the originating send event is unconstrained, however, because
these times likely come from distinct physical clocks.

%%%%%%%%%%%
\subsection{Consistency}

\begin{definition}\label{def:consistency}
For each write event on process $j$ with tag $g_j$,
let $g_i$ be the tag of the corresponding accept event on process $i$
or $\infty$ if there is no corresponding accept event.
The \textbf{inconsistency} $\bar{C}_{ij} \in \mathbb{I}$ from $j$ to $i$ is defined to be
\begin{equation}
\bar{C}_{ij} = \max(\mathcal{T}(g_i) - \mathcal{T}(g_j)),
\end{equation}
where the maximization is over all write events on process $j$.
If there are no write events on $j$, then we define $\bar{C}_{ij} = 0$.
\end{definition}

It is clear that $\bar{C}_{ij} \ge 0$.
If $\bar{C} = 0$,
we have \textbf{strong consistency}.
We will see that this strong consistency comes at a price in availability,
and that network failures can result in unbounded unavailability.
If $\bar{C}$ is finite, we have \textbf{eventual consistency}, and $\bar{C}$ quantifies ``eventual.''

Notice that inconsistency measures the difference between two \emph{logical times}.
We will show how \lf enables manipulation of the tags of events to relax consistency
requirements in order to gain availability.
It does so without sacrificing determinacy.

%%%%%%%%%%%
\subsection{Availability}

In database systems,
unavailability, $\bar{A}$, is a measure of the time it takes for a system to respond to user requests~\cite{kleppmann2015critique}.
A user request is an external event that originates from outside the distributed system.
We generalize the external events to include sensor inputs, not just user requests, and the responses
to include actuations.

Assume that a user request or sensor input triggers a read event in process $i$ with tag $g_i$ such that its timestamp $\mathcal{T}(g_i)$ is
the reading of a local clock when the external event occurs.
Let $T_i$ be the physical time of the read event, i.e., the physical time at which the read is processed.
Hence, $T_i \ge \mathcal{T}(g_i)$.

\begin{definition}\label{def:availability}
For each read event on process $i$, let $g_i$ be its tag and $T_i$ be the physical time
at which it is processed. The \textbf{unavailability} $\bar{A}_i \in \mathbb{I}$ at process $i$ is defined to be
\begin{equation}
\bar{A}_i = \max(T_i - \mathcal{T}(g_i)),
\end{equation}
where the maximization is over all read events on process $i$ that are triggered by user requests.
If there are no such read events on process $i$, then $\bar{A}_i = 0$.
\end{definition}

Because we are considering only read events that are triggered from outside the software system,
$\mathcal{T}(g_i) \le T_i$, so $\bar{A}_i \ge 0$.
If $\bar{A}_i = 0$, then we have maximum availability  (minimum unavailability).
This situation arises when external triggers cause immediate reactions.

%%%%%%%%%%%%%%%%
\subsection{Processing Offsets}

We require that each process handle events in tag order.
This gives the overall program a formal property known as \textbf{causal consistency},
which is analyzed in depth by Schwartz and Mattern.
They define a causality relation, written $e_1 \rightarrow e_2$, between events $e_1$ and $e_2$ to mean that $e_1$ can causally affect $e_2$.
The phrase ``causally affect'' is rather difficult to pin down (see Lee \cite[Chapter 11]{Lee:20:Coevolution} for the subtleties around the notion of causation),
but, intuitively, $e_1 \rightarrow e_2$ means $e_2$ cannot behave as if $e_1$ had not occurred.
Put another way, if the effect of an event is reflected in the state of a local replica of a variable $x$, then any cause of the event
must also be reflected. Put yet another way, an
observer must never observe an effect before its cause.

% Per Shaokai, the following is problematic because the causal relation is about traces, not about possible traces.
%Without more constraints, however, it can be difficult to determine what the value sent by $e_2$ might be.
%For example, if there is some other event $e_3$ that writes $x = 2$ and is not causally related to $e_1$ or $e_2$,
%then $e_3$ may send $x = 1$ or $x = 2$ even though $e_1 \rightarrow e_2$.

Formally, the causality relation of Schwartz and Mattern is the smallest transitive relation such that
$e_1 \rightarrow e_2$ if $e_1$ precedes $e_2$ in a process,
or $e_1$ is the sending of a value in one process and $e_2$ is the acceptance of the value in another process.
If neither $e_1 \rightarrow e_2$ nor $e_2 \rightarrow e_1$ holds, then we write $e_1 || e_2$ or $e_2 || e_1$ and say that $e_1$ and $e_2$ are \textbf{incomparable}.
The causality relation is identical to the ``happened before'' relation of Lamport \cite{Lamport:78:Time},
but Schwartz and Mattern prefer the term ``causality relation''
because even if $e_1$ occurs unambiguously earlier than $e_2$ in physical time,
they may nevertheless be incomparable, $e_1 || e_2$.

The causality relation is a strict partial order.
Schwartz and Mattern use their causality relation to define a ``consistent global snapshot'' of a distributed computation
to be a subset $S$ of all the events $E$ in the execution that is a downset,
meaning that if $e' \in S$ and $e \rightarrow e'$, then $e \in S$
(this was previously called a ``consistent cut'' by Mattern~\cite{Mattern:88:virtualtime}).

To maintain causal consistency, the requirement that a process have nondecreasing tags
means that, in a trace, a read or write event triggered by an external input
may have a physical time $T$ that is significantly larger than its tag's timestamp $\mathcal{T}(g)$.
While $\mathcal{T}(g)$ is determined by the physical clock at the time the external input appears,
the physical time at which the event is actually processed may have to be later to ensure that all events with earlier tags have been processed.
This motivates the following definition:

%%%%%%%%%%%%
\begin{definition}\label{def:processing}
For process $i$, the \textbf{processing offset} $O_i \in \mathbb{I}$ is
\begin{equation}
O_i = \max(T_i - \mathcal{T}(g_i))
\end{equation}
where $T_i$ and $g_i$ are the physical time and tag, respectively,
of a write event on process $i$ that is triggered by a local external input
(and hence assigned a timestamp drawn from the local clock).
The maximization is over all such write events in process $i$.
If there are no such write events, then $O_i = 0$.
\end{definition}

The processing offset closely resembles the unavailability of Definition~2,
but the former refers to \emph{write} events and the latter refers to \emph{read} events.
The processing offset, by definition, is greater than or equal to zero.

%%%%%%%%%%%
\subsection{Apparent Latency}

When a write to a shared variable occurs in process $j$, some time will elapse before a corresponding accept event
on process $i$ triggers a write to its local copy of the shared variable.
This motivates the following definition:
%%%%%%
\begin{definition}\label{def:apparent}
Let $g_j$ be the tag of a write event in process $j$ that is triggered by an external input at $j$
(so $\mathcal{T}(g_j)$ is the physical time of that external input).
Let $T_i$ be the physical time of the corresponding accept event in process $i$ (or $\infty$ if there is no such event).
(If $i=j$, we assume $T_i$ is the same as the physical time of the write event.)
The \textbf{apparent latency} or just \textbf{latency} $\mathcal{L}_{ij} \in \mathbb{I}$ for communication from $j$ to $i$ is
\begin{equation}\label{eq:apparent}
\mathcal{L}_{ij} = \max(T_i - \mathcal{T}(g_j)),
\end{equation}
where maximization is over all such write events in process $j$.
If there are no such write events, then $\mathcal{L}_{ij} = 0$.
\end{definition}

Note that $T_i$ and $\mathcal{T}(g_j)$ are physical times \emph{on two different clocks} if $i \neq j$,
so this apparent latency is an actual latency only if those clocks are perfectly synchronized.
Unless the two processes are actually using the same physical clock, they will never be perfectly synchronized.
Hence, the apparent latency may even be negative.
Note that despite these numbers coming from different clocks, if tags are sent along with messages, this apparent latency is measurable.

The apparent latency is a sum of four components,
\begin{equation}\label{eq:latency}
\mathcal{L}_{ij} = O_j + X_{ij} + L_{ij} + E_{ij} ,
\end{equation}
where $X_{ij}$ is \textbf{execution time} overhead at node $j$ for sending a message to node $i$,
$L_{ij}$ is the \textbf{network latency} from $j$ to $i$, and $E_{ij}$ is the \textbf{clock synchronization error}.
The three latter quantities are indistinguishable and always appear summed together,
so there is no point in breaking apparent latency down in this way.
Moreover, these latter three quantities would have to be measured with some physical clock,
and it is not clear what clock to use.
The apparent latency requires no problematic measurement since it explicitly refers to local clocks and tags.

The clock synchronization error can be positive or negative, whereas $O_j$, $X_{ij}$, and $L_{ij}$ are always nonnegative.
If $E_{ij}$ is a sufficiently large negative number, the apparent latency will itself also be negative.
Because of the use of local clocks, the accept event will appear to have occurred before the user input that triggered it.
This possibility is unavoidable with imperfect clocks.

When $i=j$, $\mathcal{L}_{ij} = O_j = O_i$.
The apparent latency at any node due to itself is just its processing offset.

%%%%%%%%%%%%%%
\subsection{The CAL Theorem}

The above definitions lead immediately to the following theorem:
\begin{theorem}\label{thm:cal}
Given a trace, the unavailability at process $i$ is, in the worst case,
\begin{equation}\label{eq:cal}
\bar{A}_i = \max \left ( O_i, \max_{j \in N} (\mathcal{L}_{ij} - \bar{C}_{ij}) \right ) ,
\end{equation}
where $O_i$ is the processing offset,
$\mathcal{L}_{ij}$ is apparent latency (which includes $O_j$), and
$\bar{C}_{ij}$ is the inconsistency.
\end{theorem}

This can be put in an elegant form using max-plus algebra~\cite{Baccelli:92:MaxPlus}.
Let $N$ be the number of processes, and define an $N \times N$ matrix $\Gamma$ such that its elements are given by
\begin{equation}\label{eq:matrix}
\Gamma_{ij} = \mathcal{L}_{ij} - \bar{C}_{ij} - O_j =
X_{ij} + L_{ij} + E_{ij} - \bar{C}_{ij}.
\end{equation}
That is, from (\ref{eq:latency}), the $i$, $j$-th entry in the matrix is an assumed bound on 
$X_{ij} + L_{ij} + E_{ij}$ (execution time, network latency, and clock synchronization error),
adjusted downwards by the specified tolerance for inconsistency $\bar{C}_{ij}$.

Let $\bm{A}$ be a column vector with elements equal to the unavailabilities $\bar{A}_i$,
and $\bm{O}$ be a column vector with elements equal to the processing offsets $O_i$.
Then the CAL theorem (\ref{eq:cal}) can be written as
\begin{equation}\label{eq:calmatrix}
\bm{A} = \bm{O} \oplus \Gamma \bm{O},
\end{equation}
where the matrix multiplication is in the max-plus algebra.
This can be rewritten as
\begin{equation}\label{eq:calmatrix2}
\bm{A} = (\bm{I} \oplus \Gamma) \bm{O},
\end{equation}
where $\bm{I}$ is the identity matrix in max-plus, which has zeros along the diagonal and $-\infty$ everywhere else.
Hence, unavailability is a simple linear function of the processing offsets,
where the function is given by a matrix
that depends on the network latencies, clock synchronization error, execution times, and specified inconsistency in a simple way.

%%%%%%%%%%%%%%%%%%%%%%%%%%%%%%%%%%%
\subsection{Evaluating Processing Offsets}\label{sec:conservative}

The processing offsets $O_i$ and $O_j$ are physical time delays incurred on nodes $i$ and $j$ before they can begin handling events.
Specifically, node $i$ can begin handling a user input (a write event) with tag $g_i$ at physical time $T_i = \mathcal{T}(g_i) + O_i$.
In the absence of any further information about a program, we can use our $\Gamma$ matrix to calculate these offsets.
However, the result is conservative because it does not use dependency information that may be present in a program
(and is present in the \lf programs we give in the next section).
A less conservative technique is explained below in Section~\ref{sec:intersection}.

First, in the current implementation of \lfshort, by default, logical time ``chases'' physical time, meaning that logical time never gets ahead of physical time.
To model this, define a zero column vector $\bm{Z}$ where every element is zero.
With this, we require at least that
$
\bm{O} \ge \bm{Z}.
$
Note that, in general, this vector could be given negative numbers or even $-\infty$, in which case the federate may be able to advance logical time ahead of physical time, but this is not currently supported in \lfshort, so we use a zero vector.
In addition, to ensure that node $i$ processes events in tag order, it is sufficient to ensure that node $i$ has received all network input
events with tags less than or equal to $g_i$ before processing any event with tag $g_i$.
With this (conservative) policy,
$
O_i \ge \max_j (\mathcal{L}_{ij}- \bar{C}_{ij}) .
$
The smallest processing offsets that satisfy these two constraints satisfy
\begin{equation}\label{eq:pessimistic}
\bm{O} = \bm{Z} \oplus \Gamma \bm{O} .
\end{equation}
This is a system of equations in the max-plus algebra.
From Baccelli, et al. \cite{Baccelli:92:MaxPlus} (Theorem 3.17),
if every cycle of the matrix $\Gamma$ has weight less than zero, then the unique solution of this equation is
\begin{equation}\label{eq:solution}
\bm{O} = \Gamma^* \bm{Z},
\end{equation}
where the \textbf{Kleene star} is
$
\Gamma^* = \bm{I} \oplus \Gamma \oplus \Gamma^2 \oplus \cdots .
$
Baccelli et al. (Theorem 3.20 \cite{Baccelli:92:MaxPlus}) show that this reduces to
\begin{equation}\label{eq:kleene}
\Gamma^* = \mathbf{I} \oplus \Gamma \oplus  \cdots \oplus \Gamma^{N-1},
\end{equation}
where $N$ is the number of processes.

The requirement that the cycle weights be less than zero is intuitive,
but overly restrictive.
It means that along any communication path from a node $i$ back to itself,
the sum of the logical delays $D_{jk}$ must exceed the sum of the execution times, network latencies, and clock synchronization errors
along the path.
This implies that we have to tolerate a non-zero inconsistency somewhere on each cycle.

In practice, programs may have zero or positive cycle means.
Theorem 3.17 of Baccelli, et al. \cite{Baccelli:92:MaxPlus} shows that if all cycle weights are non-positive,
then there is a solution, but the solution may not be unique.
If there are cycles with positive cycle weights, there is no finite solution for $\bm{O}$ in (\ref{eq:pessimistic}).
In this case, the only solution to (\ref{eq:pessimistic}) sets
all the processing offsets to $\infty$.
Every node must wait forever before handling any user input.
This is, of course, the ultimate price in availability.

Equation (\ref{eq:solution}) is conservative because,
absent more information about the application logic, we must assume that any network input at node $i$
with tag $g_i$ can causally affect any network output with tag $g_i$ or larger.
For particular applications, it is possible to use the detailed structure of the \lf program to derive
processing offsets that are not conservative, as we do below in Section~\ref{sec:intersection}.

%%%%%%%%%%%%%%%%%%%%%%%%%%%%%%%%%%%
\section{Availability and Consistency in \lfshort} \label{sec:lf}

In this section, we briefly introduce \lf and show how it expresses consistency
and availability requirements.
We then discuss how processing offsets can be determined.

%%%%%%%%%%%%%%%%%%%%%%%%%%%%%%%%%%%
\subsection{Brief Introduction to Lingua Franca}

\lf (or \lfshort, for short)~\cite{LohstrohEtAl:21:Towards} is 
a polyglot coordination language that orchestrates
concurrent and distributed programs written in any of several target languages
(as of this writing, C, C++, Python, TypeScript, and Rust).
In \lfshort, applications are defined as concurrent compositions of components called
\textbf{reactors}~\cite{Lohstroh:2019:CyPhy,Lohstroh:EECS-2020-235}.
\lfshort borrows the best semantic features of established models of computation, such as
actors~\cite{Agha:97:Actors}, logical execution time (LET)~\cite{Kirsch:12:LET}, synchronous reactive
languages~\cite{Benveniste:91:Synchronous}, and discrete event
systems~\cite{LeeEtAl:7:DiscreteEvents} including DEVS~\cite{Zeigler:1997:DEVS}
and SystemC~\cite{Liao:97:Scenic}.
\lfshort programs are concurrent and deterministic~\cite{Lee:21:Determinism} (except when fault handlers are invoked).
Given any set of tagged input events, there is exactly one correct behavior.

\begin{figure}[t!]

    \begin{lstlisting}[language=LF,escapechar=|]
target C; |\label{ln:target}|
reactor Sense { |\label{ln:class}|
  output out:int;    |\label{ln:output}|
  timer t(0, 10 ms); |\label{ln:timer}|
  state my_state:int(0);   |\label{ln:state}|
  reaction(t) -> out {= |\label{ln:reaction}|
    // code in C: produce out, update my_state
  =}
}
reactor Actuate {
  input in:int;    |\label{ln:input}|
  reaction(in) {= 
    // code in C: read in
  =} deadline(10 ms) {= |\label{ln:deadline}|
    // code in C: handle deadline violations|\label{ln:handler}|
  =}
}
reactor Compute {
  input in:int;
  output out:int;
  reaction(in) -> out {=  
    // code in C: read in, write out
  =}
}
main reactor { |\label{ln:main}|
  i1 = new Sense(); |\label{ln:instance}|
  i2 = new Compute();
  i3 = new Actuate();
  i1.out -> i2.in after 10 ms; |\label{ln:connection}|
  i2.out -> i3.in after 10 ms; 
}
    \end{lstlisting}
    \includegraphics[width=\columnwidth]{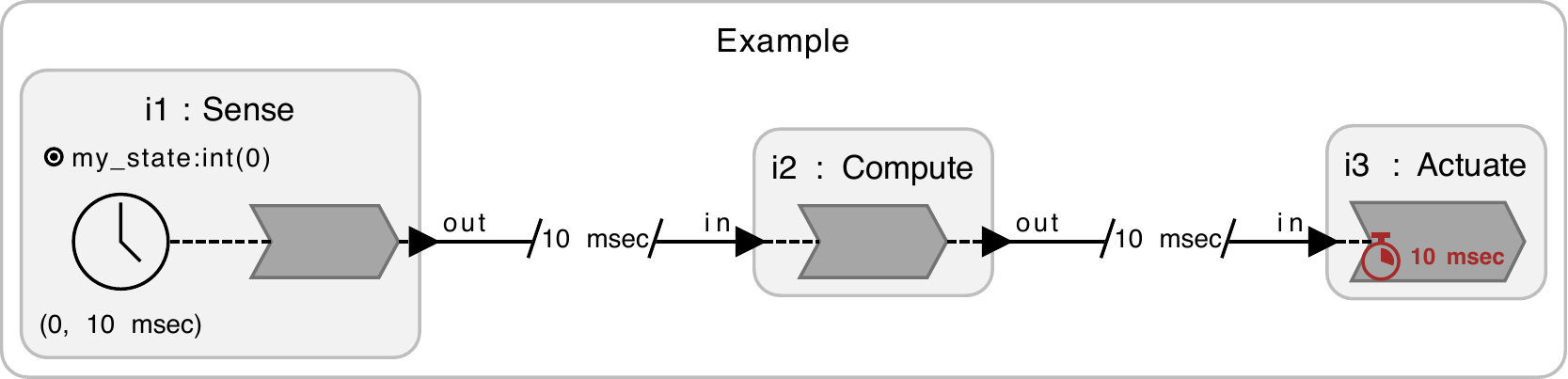}
    \caption{Structure of an \lfshort program for a simple pipeline. \label{fig:lf}}
\end{figure}

Fig.~\ref{fig:lf} gives a simple example that we use to explain the structure of an \lfshort program
and how it specifies availability and consistency requirements.
This program defines a simple pipeline consisting of a data source, a data processor,
and a data sink. The data source could, for example, poll a sensor and filter its readings.
The data processor could use the sensor data to calculate a command to send to an actuator.
The data sink could drive the actuator.

The diagram at the bottom of the figure is automatically generated by the \lfshort tools given the
source code at the top.\footnote{The diagram synthesis feature was created
by Alexander Schulz-Rosengarten of Kiel University using the graphical layout tools from the
KIELER Lightweight
Diagrams framework~\cite{SchneiderSvH13} (see \url{https://rtsys.informatik.uni-kiel.de/kieler}).}
In later examples, we will show the diagram only and not the source code
because the diagram contains sufficient information.
The chevrons in the figure represent \textbf{reactions}, which process events, and their dependencies on inputs and their ability to produce outputs
is shown using dashed lines.

Line \ref{ln:target} in the source code defines the target language, which is the language in which reactions are
written, and the language into which the entire \lfshort program is translated.
This example uses the C target, which means that the bodies of reactions are written in C.

Line \ref{ln:class} declares a \textbf{reactor class} named \texttt{Sense}, which has
an \textbf{output port} (line \ref{ln:output}),
a \textbf{timer} (line \ref{ln:timer}),
a \textbf{state variable} (line \ref{ln:state}),
and a \textbf{reaction} (line \ref{ln:reaction}).
The output port has name \texttt{out} and type \texttt{int}.
The timer has name \texttt{t}, offset 0 (meaning it should start when the program starts), and period 10 ms.
The state variable has a name, type, and initial value.
Each of these properties of the reactor is represented in the diagram at the bottom.

A \textbf{reaction}, like that on line \ref{ln:reaction}, is defined with a syntax of the form
\begin{quote}
\textbf{reaction}($L_1$) $L_2$ \texttt{->} $L_3$ \{= code body =\}
\end{quote}
where $L_1$ is a list of \textbf{triggers}, which are inputs, timers, and actions (we will discuss actions later);
$L_2$ is an optional list of \textbf{observables}, which are inputs and actions that do not trigger the reaction but may be read by the reaction; and
$L_3$ is an optional list of \textbf{effects}, which are outputs, actions, and modes (which are not used in this paper).

The particular reaction on line \ref{ln:reaction} is triggered by the timer every 10 ms.
When it is triggered for the $n$-th time, its logical time will be $t = s + n \times 10$ ms, where $s$ is the logical
start time, typically set using the local physical clock when the program starts.
The runtime system attempts to align logical time with physical time, so this reaction will be
invoked roughly every 10 ms, but this cannot be done perfectly.
By default, logical time ``chases'' physical time
in a program execution, so that reactions with a logical time $t$ are invoked
close to (but never before) physical time $T = t$.

A reaction may optionally have a \textbf{deadline}, as shown in the \texttt{Actuate} reactor class on line \ref{ln:deadline}.
This gives a time value
and a code body to execute instead of the reaction if the deadline is violated.
The time value may be a parameter of the reactor class but here is shown as the constant 10 ms.
A deadline with time value $d = 10$ms is violated for an event with tag $g$ if the reaction is invoked at
a physical time $T$ where $T > \mathcal{T}(g) + d$.
Such a deadline explicitly specifies an availability requirement.
The deadline violation handler (line \ref{ln:handler}) is a fault handler.
The \lfshort runtime system uses an EDF scheduler to attempt to avoid violating this deadline,
as usual in real-time systems.

The top-level (main) reactor is defined on line \ref{ln:main}.
Within it, reactor instances are created as on line \ref{ln:instance} using the \textbf{new} keyword.
If the \textbf{main} keyword is replaced with \textbf{federated},
then a separate C program is generated for each reactor instantiated within the federated reactor.
Otherwise, a single multi-threaded C program is generated for the entire program.
For the federated case, each instance is called a \textbf{federate}, and
tagged inputs will arrive from the network at the input ports and be handled in tag order.

The routing of messages is specified by \textbf{connections}, as shown on line \ref{ln:connection},
which connects the output of \texttt{i1} to the input of \texttt{i2}.
Such a connection may optionally alter the timestamp of the message using the \textbf{after} keyword.
The connection on line \ref{ln:connection} specifies that the timestamp of the input event at \texttt{i2}
should be 10 ms greater than  the timestamp at the output of \texttt{i1}.
Such a \textbf{logical delay} explicitly relaxes the consistency requirements because it explicitly states
that \texttt{i2} can use information that is 10 ms out of date relative to \texttt{i1}.

We can see immediately that use of logical delays improves availability for this example,
as expected from the CAL theorem.
Suppose that this program is federated and that the three instances are mapped to distinct processors on
a network.
Were it not for the logical delays, intuitively, the \texttt{Actuate} reactor \texttt{i3} would be unable to
react until the message from the \texttt{Sense} reactor \texttt{i1} had flowed through the network to
\texttt{i2}, \texttt{i2} had completed its reaction, and the result from \texttt{i2} had flowed over the network
to \texttt{i3}.
If these delays add up to more than 10 ms, the \texttt{i3}'s deadline will be missed.
With the logical delays, however, as long as the delays add up to less than 30 ms, the deadline
will not be missed.
If the delays add up to less than 20 ms, then \texttt{i3} can react to its input with timestamp $t$ as soon
as physical time $T$ matches or exceeds $t$.
The specified tolerance for inconsistency improves availability.

This intuition can be made rigorous using the CAL theorem.

%%%%%%%%%%%%%%%%%%%%%%%%%%%%%%%%%%%%%%%%
\subsection{Evaluating Unavailability}

For the program in Fig.~\ref{fig:lf}, the $\Gamma$ matrix is given by
\begin{equation}\label{eq:gamma}
\Gamma = \left [
    \begin{matrix}
    0 & -\infty & -\infty \\
    \Gamma_{21} & 0 & -\infty \\
    -\infty & \Gamma_{32} & 0
    \end{matrix}
    \right ]
\end{equation}
where 
\begin{itemize}
\item $\Gamma_{21} = X_{21}+L_{21}+E_{21} - 10 \mbox{ms}$,
\item $\Gamma_{32} = X_{32}+L_{32}+E_{32} - 10 \mbox{ms}$,
\end{itemize}
and
\begin{itemize}
    \item $X_{21}$ is the execution time for the reaction in \texttt{Sense},
    \item $L_{21}$ is the network latency from \texttt{Sense} to \texttt{Compute},
    \item $E_{21}$ is the clock synchronization error from \texttt{i1} to \texttt{i2},
    \item $X_{32}$ is the execution time for the reaction in \texttt{Compute},
    \item $L_{32}$ is the network latency from \texttt{Compute} to \texttt{Actuate}, and
    \item $E_{32}$ is the clock synchronization error from  \texttt{i2} to \texttt{i3}.
\end{itemize}
The $-\infty$ entries in the matrix are a consequence of a lack of communication.

On the communication path from \texttt{i1} to \texttt{i2}, there is a logical delay of 10 ms, which is an explicit declaration of an inconsistency $\bar{C}_{21} = 10$ ms.
The \texttt{Compute} reactor's view of the data from the \texttt{Sense} reactor is 10 ms behind.
We can now determine that this allowance of 10 ms of inconsistency improves availability compared to what we would get without it.

First, we can use the analysis of Section~\ref{sec:conservative} to evaluate the processing offsets.
For this example, $N = 3$, so (\ref{eq:kleene}) reduces to
\[
\Gamma^* = \mathbf{I} \oplus \Gamma \oplus  \Gamma^2.
\]
It is straightforward to evaluate this to get
\[
\Gamma^*  = \left [
    \begin{matrix}
    0 & -\infty & -\infty \\
    \Gamma_{21} & 0 & -\infty \\
    \Gamma_{21} + \Gamma_{32} & \Gamma_{32} & 0
    \end{matrix}
    \right ]
\]
Intuitively, this matrix captures the fact that the \texttt{Actuate} reactor indirectly depends on
the \texttt{Sense} reactor, something not directly represented in the $\Gamma$ matrix.

We can now evaluate (\ref{eq:solution}) to get
\begin{equation}\label{eq:po}
\bm{O} = \Gamma^* \bm{Z} = \left [
    \begin{matrix}
    0 \\
    \max( \Gamma_{21}, 0 ) \\
    \max(\Gamma_{21} + \Gamma_{32}, \Gamma_{32}, 0)
    \end{matrix}
    \right ]
\end{equation}
Next, we evaluate (\ref{eq:calmatrix2}) to get the unavailability at each node,
\begin{equation}\label{eq:unav}
\bm{A} = (\bm{I} \oplus \Gamma) \bm{O} = \left [
    \begin{matrix}
    0 \\
    \max( \Gamma_{21}, 0 ) \\
    \max(\Gamma_{21} + \Gamma_{32}, \Gamma_{32}, 0)
    \end{matrix}
    \right ]
\end{equation}
In this simple case, the unavailability is equal to the processing offsets,
which means that the processing offsets capture all the waiting that needs to be done
to realize the semantics of the program.

These unavailability numbers are intuitive.
First, note that the \texttt{Sense} can react to external stimulus immediately.
It has no network inputs to worry about, so $\bm{A}_0 = 0$.
The \texttt{Compute} reactor, however, can react to an input stimulus with timestamp $t$ immediately
when physical time $T=t$ only if $X_{21} + L_{21} + E_{21} \leq 10 \mbox{ms}$.
Otherwise, in order to ensure that it processes events in timestamp order, it may need to wait
until $T = t + X_{21} + L_{21} + E_{21} - 10 \mbox{ms}$.
Similarly, the \texttt{Actuate} reactor can respond immediately if
$\Gamma_{32} \leq 0$ and $\Gamma_{21} + \Gamma_{32} \leq 0$.
These conditions occur if the 10 ms logical delay is larger than the apparent latencies in communication.

If we change line \ref{ln:connection} to this subtly different version:
\begin{lstlisting}[firstnumber=29,language=LF,escapechar=|]
  i1.out |\physicalConn| i2.in;
\end{lstlisting}
then there is no upper bound on the inconsistency between these two instances.
The subtle change is to replace the \textbf{logical connection} \texttt{->} with
a \textbf{physical connection} \physicalConn.
In \lf, this is a directive to assign a new tag $g_i$ at the receiving end $i$
based on a local measurement of physical time $T_i$ when the message is received such that $\mathcal{T}(g_i) = T_i$.
The original tag is discarded.
Such connections, therefore, have no effect on availability,
but they completely abandon consistency.

%%%%%%%%%%%%%%%%%%%%%%%%%%%%%%%%%%%
\subsection{Processing Offsets in \lf}

\lf offers two coordination strategies for federated execution, \textit{centralized} and \textit{decentralized}~\cite{BateniEtAl:22:Xronos}.
The centralized coordinator is an extension of the
High-Level Architecture (HLA)~\cite{dahmann1997department}, a distributed discrete-event simulation
standard.
The decentralized coordinator is an extension of
PTIDES~\cite{Zhao:07:PTIDES}, a real-time technique also implemented in Google
Spanner~\cite{CorbettEtAl:12:Spanner}, a globally distributed database. 
This coordinator is also influenced by Lamport~\cite{Lamport:84:TimeStamps} and Chandy and
Misra~\cite{ChandyMisra:79:DDE,Chandy:86:DDE}.

For the purposes of this paper, we only need to know how these coordination mechanisms relate to processing
offsets and availability. The \textit{centralized} coordinator is the easiest to understand.
It does whatever is necessary to ensure that events are processed in tag order.
In particular, execution in a federate will be delayed when such a delay is needed to ensure tag order semantics.
As a consequence, the processing offsets and unavailability bounds emerge from an execution of the program.
As network latencies vary, the offsets and unavailability vary.
The CAL theorem tells what to expect these numbers to be, given network latencies.
The programmer, therefore, can use the CAL theorem to determine whether deadlines will be met,
given specific latencies.

The processing offsets and unavailability bounds play a bigger role when using the \textit{decentralized} coordinator.
In particular, with this coordinator, the programmer is required to specify a \textbf{safe-to-advance} (\textbf{STA})
offset for each federate. The choice of STA at federate $i$ can be guided by processing offset $\bm{O}_i$ for federate $i$, with the
caveat that $\bm{O}_i$ is a property of a \emph{trace}, whereas STA is a property of a \emph{program} (a family of traces).
The STA specified for a federate $i$ \textbf{enforces} $O_i$ during execution for all the traces produced by federate $i$.
For particular reactions, the programmer can also give an additional \textbf{safe-to-assume-absent} (\textbf{STAA}) offset.
STAA gives an additional time beyond the STA to wait before assuming that the absence of a message on a particular input means that there will be no message on that port with the current tag or less.
The availability bound for any shared state read by that reaction can similarly guide the choice of STAA.
Such guidance is much better than guesswork.

Both coordinators will deterministically execute the \lfshort program identically, yielding the same behavior
as an unfederated execution, under certain assumptions.
For the centralized coordinator, the assumptions are that the network latencies are sufficiently low that
no deadlines are missed.
For the decentralized coordinator, the assumptions are that the network latencies are sufficiently low that
events are seen by each reactor in tag order.
In both cases, violations of the assumptions are detectable and can be handled by fault-handling code provided
by the programmer.

The key difference between the two coordinators, therefore, is in their fault handling.
When network latencies get large (or the network gets partitioned), the centralized coordinator
sacrifices availability, whereas the decentralized coordinator sacrifices consistency.
Which of these is the right choice is application dependent.  

Note also that for both coordinators, the CAL theorem can be used to derive the requirements
on network latencies, and therefore provides a principled guide for choosing
networking technology and can guide refactoring designs to move computations
between embedded, edge, and cloud computing.

%%%%%%%%%%%%%%%%%%%%%%%%%%%%%%%%%%%
\subsection{Fault Handling}

Any assumptions about network latency may be violated in the field. In centralized coordination, such violations will manifest as deadline violations, whereas in decentralized coordination, they will manifest as consistency violations. In both cases, \lfshort allows the programmer to specify exception handlers to be invoked when such violations occur.

Hence, for safety-critical CPS applications, the proposed framework promises some astonishingly attractive properties. First, a programmer can explicitly decide when and how much to give up consistency and when and how much to give up availability to accommodate execution times, network latency, and clock synchronization error. 
Second, the programmer's specification will imply explicit constraints on the technology (networking, processing, and clock synchronization) that can be used to guide selection of parts to use and mapping of software components onto resources (embedded, edge, or cloud).
Third, to allow for (inevitable) possible failures in the field, where specifications are not met due to unforeseen circumstances such as hardware failures, the programmer can explicitly give code to execute when the fault occurs.
%There is no software framework currently that has these properties.

We will show next how these principles can be applied through two complementary practical examples.

%%%%%%%%%%%%%%%%%%%%%%%%%%%%%%%%%%%
\section{Tradeoffs in Practical Systems} \label{sec:tradeoffs}

In this section, we consider two safety-critical real-time systems with complementary properties.
The first demands that availability (timely responses) be prioritized over consistency in the presence of faults.
This example also requires a measured relaxation of consistency to meet tight deadlines.
The second example demands that consistency be prioritized over availability in the presence of faults.
The second example also illustrates how cycles are handled by the CAL theorem.

%%%%%%%%%%%%%%%%%%%%%%%%%%%%%%%%%%%
\begin{figure*}[!t]
\centering
\includegraphics[width=1.4\columnwidth]{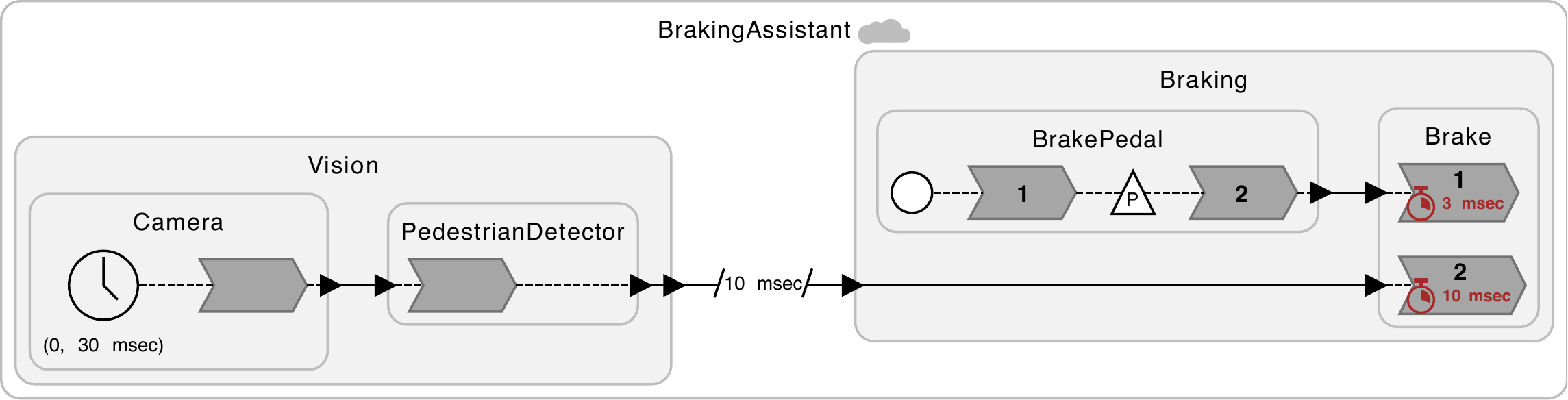}
\includegraphics[width=0.6\columnwidth]{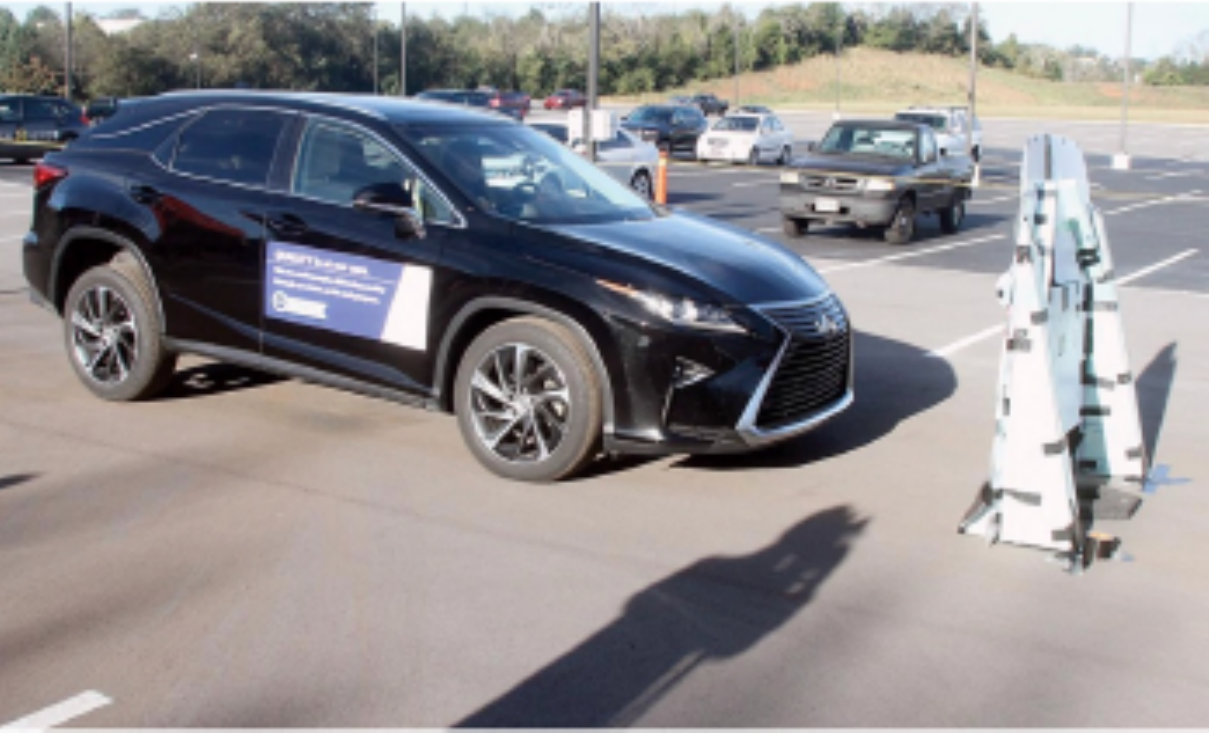}
\caption{ADAS system schematic and photo from 2018 demo by Denso, reported in \textit{The Daily Times}.}
\label{fig:adas}
\end{figure*}

\subsection{ADAS} \label{sec:adas}

Consider an Advanced Driver Assist Systems (ADAS), as shown in Fig.~\ref{fig:adas}.
Such a system uses a camera with a computer vision system that analyzes images for pedestrians and applies braking when a pedestrian is detected, as shown in the photo.
The figure shows the diagram synthesized from an \lfshort program that shows the structure of this system.
In this structure, there are two federated reactors, a \texttt{Vision} subsystem and a \texttt{Braking} subsystem.
The structure is a pipeline similar to that of Fig.~\ref{fig:lf}, except with a twist.
Inside the \texttt{Braking} subsystem is a second sensor, which senses when a driver presses on the brake pedal.
Both the camera and the pedal can affect the same actuator, which is driven by the \texttt{Brake} reactor.

The \texttt{Vision} federate has a time-triggered periodically invoked reaction that captures and analyzes an image.
It then sends the results over the network to the \texttt{Braking} federate.
The \texttt{Braking} federate has a local interface to a sensor on the brake pedal, represented in the diagram by the triangle with a ``P'' (which represents a \textbf{physical action} in \lfshort).
When the brake pedal is pushed, an event is generated that is assigned a time stamp from the local clock
and triggers invocation of the reaction labeled ``2'' within the \texttt{BrakePedal} reactor.

Let the \texttt{Vision} federate be process 1 and the \texttt{Braking} federate be process 2. Then the $\Gamma$ matrix is similar to (\ref{eq:gamma}):
\begin{equation}
\Gamma = \left [
    \begin{matrix}
    0 & -\infty \\
    \Gamma_{21} & 0
    \end{matrix}
    \right ]
\end{equation}
where
\begin{itemize}
\item $\Gamma_{21} = X_{21}+L_{21}+E_{21} - 10 \mbox{ms},$
\end{itemize}
and
\begin{itemize}
    \item $X_{21}$ is the execution time in \texttt{Vision} to prepare the data to send to \texttt{Braking},
    \item $L_{21}$ is the network latency from \texttt{Vision} to \texttt{Braking}, and
    \item $E_{21}$ is the clock synchronization error.
\end{itemize}
The logical delay of 10 ms on the communication path from 1 to 2 is an explicit declaration of an inconsistency $\bar{C}_{21} = 10$ ms.
The Braking system's view of the sensor data from the Vision system is 10 ms behind.
Using the same methods, the processing offsets and unavailability are similar to (\ref{eq:po}) and (\ref{eq:unav}):
\[
\bm{O} = \left [
    \begin{matrix}
    0 \\
    \max( \Gamma_{21}, 0 )
    \end{matrix}
    \right ]
\quad
\bm{A} = \left [
    \begin{matrix}
    0 \\
    \max( \Gamma_{21}, 0 )
    \end{matrix}
    \right ]
\]
The allowance of 10 ms of inconsistency improves availability compared to what we would get without it.
In particular, if $X_{21}+L_{21}+E_{21} \leq 10 \mbox{ms},$ then the processing offsets and unavailability are
all zero.

In Fig.~\ref{fig:adas}, notice that the first reaction of \texttt{Brake} has a deadline of 3 ms.
This deadline is as an explicit requirement for availability, stating, effectively, that we require
\[
    X_{21} + L_{21} + E_{21} - 10 \mbox{ms} + X_{22} \le 3\mbox{ms},
\]
where $X_{22}$ is the execution time of reaction 2 in \texttt{BrakePedal}.\footnote{Note that \lfshort implements an EDF scheduling policy, and that deadlines are inherited upstream, so
reaction 2 of \texttt{BrakePedal} will have high priority.}
The requirement becomes
\begin{equation}
    X_{21} + L_{21} + E_{21} < 13\mbox{ms} - X_{22}.
\end{equation}
This requirement almost certainly means that the vision processing cannot be done in the cloud.
If it is, then the deadline is likely to be violated.
In principle, this analysis can be automated, so that a system designer simply enters the requirements (by specifying deadlines, communication paths, and consistency requirements), and the system provides feedback on the realizability of the requirements.

If the system designer really wants to do the vision processing in the cloud, then these results can be used to negotiate a service-level agreement, for example, with the 5G network vendor and the cloud service provider.
Alternatively, the 10ms tolerance for inconsistency could be increased, although this would require an evaluation of whether the ADAS system continues to be able to do its job safely.

Once the requirements and assumptions are specified, then the next key decision is what to do when
those assumptions are violated.
For this application, missing the deadline could be disastrous, so we should emphasize availability over consistency.
To accomplish that in \lf, we just have to specify to use decentralized coordination.
With this coordination mechanism, if messages fail to arrive on time from the network,
each local runtime system assumes there are no messages and continues accordingly.
This will ensure that the brake pedal event gets handled as long as the \texttt{Braking} federate's host computer
is still working.

%%%%%%%%%%%%%%%%%%%%%%%%%%%%%%%%%%%
\subsection{Four-Way Intersection} \label{sec:intersection}

\begin{figure*}[!t]
\centering
\includegraphics[width=1.15\columnwidth]{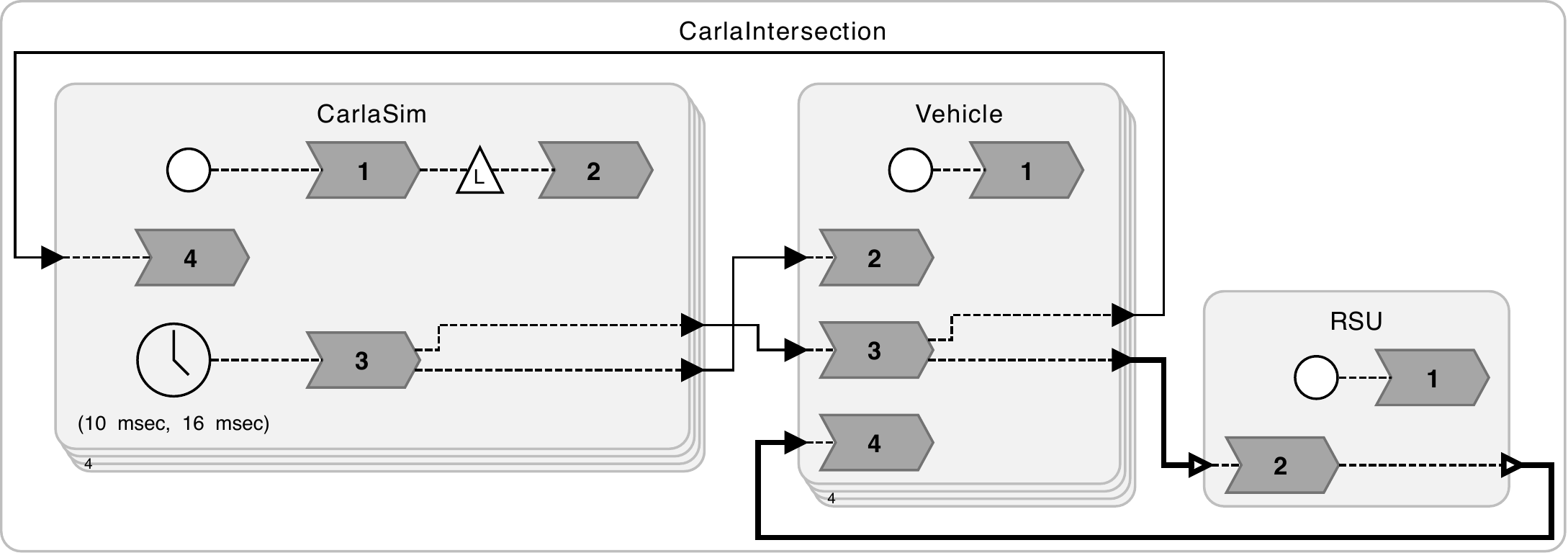}
\includegraphics[width=0.35\columnwidth]{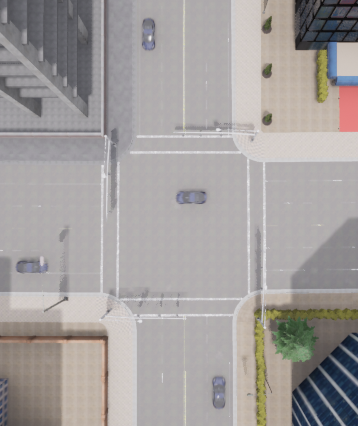}
\caption{Four-way intersection example.}
\label{fig:intersection}
\end{figure*}

Consider autonomous or semi-autonomous vehicles that leverage communication with a roadside unit (RSU) to mediate access to a four way intersection.
There are many projects working on such automation to improve traffic flow~\cite{RAO199449,Sepasgozar:22:NetworkTraffic}.
A prototype is depicted in Fig.~\ref{fig:intersection}.
This prototype uses a popular open-source vehicle simulator called Carla, which generates the animated image in the figure that gets updated as the program runs.
The prototype is implemented using the Python target in \lfshort, which enables easy integration of large legacy subsystems, such as Carla, that have Python APIs.

The \lfshort program depicted in Fig.~\ref{fig:intersection} consists of nine top-level components,
four vehicle simulators, four vehicle controllers,
and one roadside unit.
The program uses a compact \lfshort notation for banks of reactors and a multiplicity of communication channels.
In this program, as a vehicle approaches the intersection, it communicates with the RSU, sending its kinematic state (position and velocity). 
The RSU handles competing requests for access to a single shared resource, the intersection, by granting time windows to particular vehicles during which they may use the intersection.
This application represents a common pattern that occurs whenever distinct agents contend for access to a shared resource.

A key property of this application is that, very much unlike the ADAS example in Section~\ref{sec:adas}, consistency is far more important.
All vehicles and the RSU \emph{must} have a consistent view of the state of the intersection before \emph{any} vehicle can enter the intersection.
In other words, we prefer that vehicles stop (making the intersection \emph{unavailable}) over having them enter the intersection without a consistent view on the state of the system, which could lead to a collision.

In \lfshort, if we choose centralized coordination for the federated execution, the system will emphasize consistency over
availability in the event of faults (violations of the assumptions and requirements).
If messages from the network do not arrive on time, each federate stops progressing,
which will prevent a vehicle from entering the intersection.

The contrast between the requirements of the intersection and ADAS examples demonstrates that tradeoffs between availability and consistency are application dependent.
System designers should be able to make such tradeoffs in system requirements,
and software needs to be designed so it responds to faults in coherence with the stated requirements.
For the ADAS example, we need to sacrifice consistency, whereas for the intersection example, we need to sacrifice availability.

A second key difference between this intersection example and the ADAS example is that the program has
communication cycles without logical delays.
This changes how we do the analysis because we will no longer be able to use
(\ref{eq:solution}) to determine the processing offsets.
Instead, we have to derive the processing offsets using more detailed information about the program structure.
We now show how to do that.

\begin{figure}[tb]
\centering
\includegraphics[width=0.6\columnwidth]{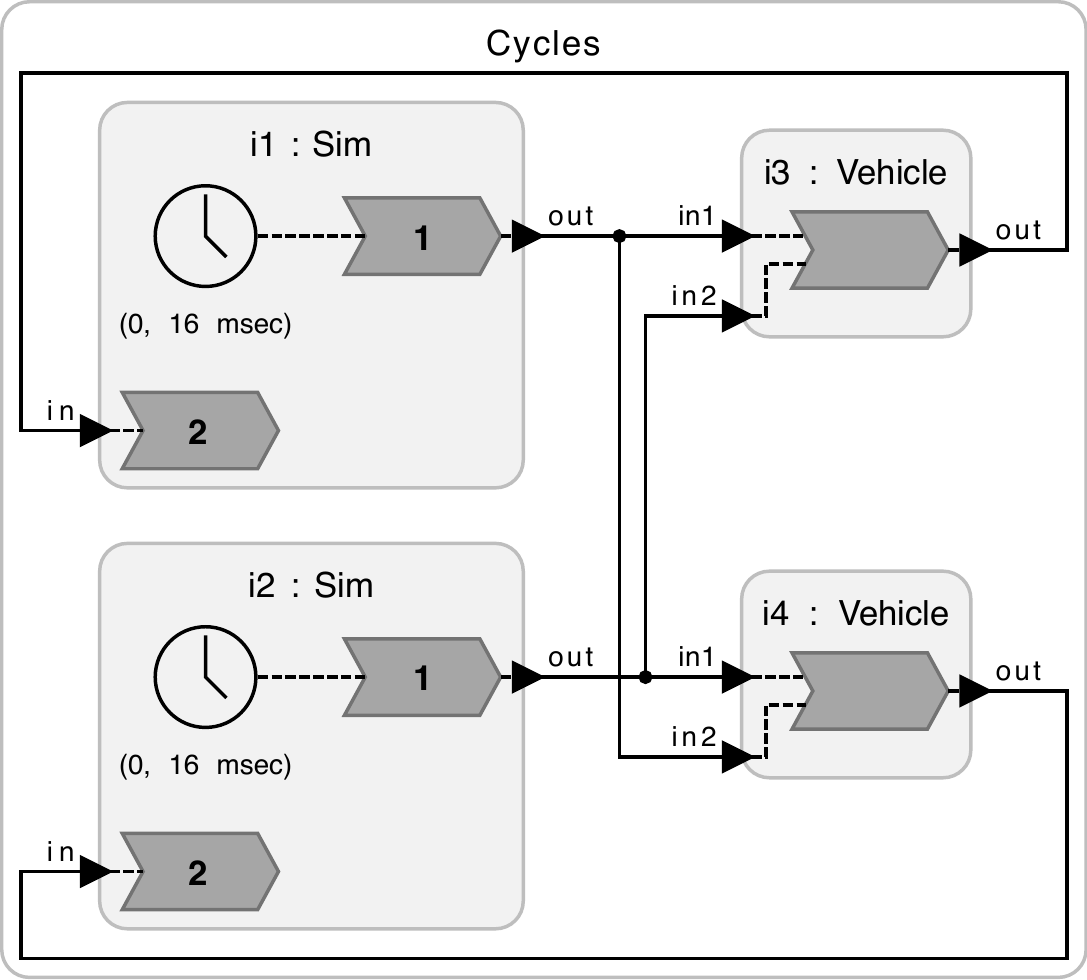}
\caption{Simplified program with cycles.}
\label{fig:cycles}
\end{figure}

The simpler \lfshort program depicted in Fig.~\ref{fig:cycles} has the essential structure of the intersection
example reduced to the minimum that illustrates the issues.  The $\Gamma$ matrix is
\[
\Gamma = \left [
    \begin{matrix}
    0 & -\infty & \Gamma_{13} & -\infty \\
    -\infty & 0 & -\infty & \Gamma_{24} \\
    \Gamma_{31} & \Gamma_{32} & 0 & -\infty \\
    \Gamma_{41} & \Gamma_{42} & -\infty & 0
    \end{matrix}
    \right ]
\]
The finite non-zero entries are defined as before,
\[
\Gamma_{ij} = X_{ij} + L_{ij} + E_{ij} - \bar{C}_{ij},
\]
where, in this case, $\bar{C}_{ij} =0$ because none of the connections has a logical delay.
To find the processing offsets, we use information that is evident in the \lfshort program but not
in the matrix.
Specifically, note that any inputs that arrive at the inputs of the \texttt{Sim} reactors will have the
same tag as an output produced by one of the two \texttt{Sim} reactors.
Moreover, the two \texttt{Sim} reactors' outputs are driven by timers with the same offset and period (zero and 16 ms),
so these outputs are logically simultaneous. We now make a key assumption:
\begin{assumption}\label{as:period}
The period of the timers is greater than any unavailability.
\end{assumption}

With this assumption, each \texttt{Sim} reactor will have completed processing all events with timestamp $t$
before it needs to advance to logical time $t + p$, where $p$ is the timer period.
Recall that we are assuming that logical time chases physical time, so physical time has to advance to $t+p$ before the federate
will even attempt to advance its logical time.
At this point, it can safely advance its logical time immediately.
Hence, the processing offset for both \texttt{Sim} reactors is zero
if our assumption \ref{as:period} is true.
  
The processing offset for the two \texttt{Vehicle} is easier to derive.
It simply depends on the communication latencies, clock synchronization errors, and
execution time bounds. The resulting processing offset vector is
\begin{equation}\label{eq:simoffset}
\bm{O} = 
\left [
    \begin{matrix}
    0 \\
    0 \\
    \max (\Gamma_{31}, \Gamma_{32}) \\
    \max (\Gamma_{41}, \Gamma_{42})
    \end{matrix}
    \right ]
\end{equation}
We can now use (\ref{eq:calmatrix2}) to calculate the unavailability:
\begin{equation}\label{eq:simavailability}
\bm{A} = (\bm{I} \oplus \Gamma) \bm{O} = \left [
    \begin{matrix}
    \Gamma_{13} + \max (\Gamma_{31}, \Gamma_{32}) \\
    \Gamma_{24} + \max (\Gamma_{41}, \Gamma_{42}) \\
    \max (\Gamma_{31}, \Gamma_{32}) \\
    \max (\Gamma_{41}, \Gamma_{42})
    \end{matrix}
    \right ]
\end{equation}
We can now see that if the clock period is less than that top two entries in this vector,
then assumption \ref{as:period} will be violated, so this becomes a requirement.

These results are intuitive and correspond with observation when we run the federated \lfshort program.
Assumption \ref{as:period} asserts that the period of the clocks is large enough that
each period begins fresh without an accumulated backlog of unprocessed events.
The execution will begin each period by advancing the logical time of each \texttt{Sim} federate to the
next period as soon as physical time matches that logical time. The zeros in the first two entries
of (\ref{eq:simoffset}) tell us this is done without delay.
The logical time of the two \texttt{Vehicle} federates, however, cannot be advanced until enough
physical time has elapsed to allow for propagation of events from \textit{both} \texttt{Sim} federates.
This delay is represented by last two entries in (\ref{eq:simoffset}).
%which will be positive unless clock
%synchronization error is large enough and of the right polarity to make them negative.

As shown in (\ref{eq:simavailability}), the unavailability of the two \texttt{Vehicle} federates
matches their processing offset. This is not surprising because they each have only one reaction
and that reaction reacts to both network inputs. However, the unavailability at the \texttt{Sim} reactors
is larger than their processing offset.
This reflects the fact that reaction 2 in each of the \texttt{Sim} reactors has to wait for the upstream
\texttt{Vehicle} reactors to execute and for their results to propagate over the network.
In other words, strong consistency---which lets actuation be logically simultaneous with the acquisition of sensor inputs---comes at the cost of a penalty in availability. The actuation is delayed in physical time, and,
more fundamentally, the period with which sensing and actuation can be done has a lower bound
that depends on the network delays.

Under centralized coordination, the actual values of all the $\Gamma_{ij}$ latencies are determined
automatically at runtime as apparent latencies.
If the program fails to keep up for any reason (e.g. network failures), then
the centralized coordinator will preserve consistency; unavailability will rise
and deadlines (if any are specified) will be missed.
Fault handlers provided by the programmer can adapt the system accordingly.
Moreover, in this case, assumption \ref{as:period} becomes invalid, so the derived unavailability bound
becomes invalid.
Unavailability will exceed our calculated bound for such a trace and, in the event of total network failure, will grow without bound.

Under decentralized coordination, the programmer chooses numbers to the $\Gamma_{ij}$ latencies based on
assumptions about network behavior and
derives processing offsets (\ref{eq:simoffset}) and unavailability (\ref{eq:simavailability}).
These then guide the choices of STA and STAA numbers specified in the program.
At runtime, each federate proceeds on the assumption that the network latencies will be respected.
If these assumptions are violated, then a reactor may see events out of timestamp order,
in which case a fault handler will be invoked.
If the network fails altogether, however, no reactor will see events out of timestamp order,
and no fault handler will be invoked.
Instead, each vehicle will act based on inconsistent information.
Hence, with decentralized coordination, availability is prioritized over consistency when a fault occurs,
which is the wrong choice for this application.

%%%%%%%%%%%%%%%%%%%%%%%%%%%%%%%%%%%%%%
\section{Conclusions}\label{sec:conclusion}

The CAL theorem, which generalizes Brewer's CAP theorem,
quantifies the relationship between inconsistency, unavailability, and apparent latency in distributed systems,
where apparent latency includes network latency, execution time overhead, and clock synchronization error.
The relationship is a linear system of equations in a max-plus algebra.
We have applied this theorem to distributed real-time systems, showing how consistency
affects the ability to bound the time it takes to react to an external stimulus, such as a sensor input,
and produce a response, an actuator output.
These bounds (which we call unavailability) depend on apparent latency and can be reduced by explicitly
relaxing consistency requirements.
Moreover, because the CAL theorem defines the effect of
network latency on the responsiveness of a system,
it can serve to guide placement of software components in end devices, in edge computers, or
in the cloud.  The consequences of such choices can be derived rather than measured or intuited.

We have shown how the \lf coordination language enables arbitrary tradeoffs between consistency and availability as apparent latency varies.
We have also shown how \lfshort programs can define fault handlers, sections of code that are executed when
specified consistency and availability requirements cannot be met because apparent latency has exceeded
the assumed bounds.
Because of its deterministic semantics, \lfshort provides predictable and repeatable behaviors in the absence of faults.
And when faults occur, \lfshort provides mechanisms for the system to adapt.

%\begin{wrapfigure}{r}{0.3\textwidth}
%  \begin{center}
%    \includegraphics[width=0.3\columnwidth]{Figures/Romi.png}
%  \end{center}
%  \caption{\lfshort running on a TI-RSLK robot controlled by an nRF52DK board combined with a Berkeley Buckler board.}
%  \label{fig:romi}
%\end{wrapfigure}

\bibliographystyle{plain}
\bibliography{Refs.bib}

\begin{thebibliography}{10}

\bibitem{Abadi:12:CAP}
Daniel Abadi.
\newblock Consistency tradeoffs in modern distributed database system design:
  {CAP} is only part of the story.
\newblock {\em Computer}, 45(2):37--42, February 2012.

\bibitem{Agha:97:Actors}
Gul~A. Agha.
\newblock Abstracting interaction patterns: A programming paradigm for open
  distributed systems.
\newblock In E.~Najm Stefani and J.-B., editors, {\em Formal Methods for Open
  Object-based Distributed Systems, IFIP Transactions}. Chapman and Hall, 1997.

\bibitem{Baccelli:92:MaxPlus}
F.~Baccelli, G.~Cohen, G.~J. Olster, and J.~P. Quadrat.
\newblock {\em Synchronization and Linearity, An Algebra for Discrete Event
  Systems}.
\newblock Wiley, New York, 1992.

\bibitem{BateniEtAl:22:Xronos}
Soroush Bateni, Marten Lohstroh, Hou~Seng Wong, Rohan Tabish, Hokeun Kim,
  Shaokai Lin, Christian Menard, Cong Liu, and Edward~A. Lee.
\newblock Xronos: Predictable coordination for safety-critical distributed
  embedded systems.
\newblock {\em arXiv:2207.09555 [cs.DC]}, July 2022.

\bibitem{Benveniste:91:Synchronous}
Albert Benveniste and G\'{e}rard Berry.
\newblock The synchronous approach to reactive and real-time systems.
\newblock {\em Proceedings of the IEEE}, 79(9):1270--1282, 1991.

\bibitem{Brewer:00:CAP}
Eric Brewer.
\newblock Towards robust distributed system.
\newblock In {\em Symposium on Principles of Distributed Computing (PODC)},
  2000.
\newblock Keynote talk.

\bibitem{Brewer:12:CAP}
Eric Brewer.
\newblock {CAP} twelve years later: How the "rules" have changed.
\newblock {\em IEEE Computer}, 45(2):23--29, February 2012.

\bibitem{Cataldo:06:Tetric}
Adam Cataldo, Edward~A. Lee, Xiaojun Liu, Eleftherios Matsikoudis, and Haiyang
  Zheng.
\newblock A constructive fixed-point theorem and the feedback semantics of
  timed systems.
\newblock In {\em Workshop on Discrete Event Systems (WODES)}, 2006.

\bibitem{ChandyMisra:79:DDE}
K.~Mani Chandy and Jayadev Misra.
\newblock Distributed simulation: A case study in design and verification of
  distributed programs.
\newblock {\em IEEE Trans. on Software Engineering}, 5(5):440--452, 1979.

\bibitem{CorbettEtAl:12:Spanner}
James~C. Corbett, Jeffrey Dean, Michael Epstein, Andrew Fikes, Christopher
  Frost, JJ~Furman, Sanjay Ghemawat, Andrey Gubarev, Christopher Heiser, Peter
  Hochschild, Wilson Hsieh, Sebastian Kanthak, Eugene Kogan, Hongyi Li,
  Alexander Lloyd, Sergey Melnik, David Mwaura, David Nagle, Sean Quinlan,
  Rajesh Rao, Lindsay Rolig, Yasushi Saito, Michal Szymaniak, Christopher
  Taylor, Ruth Wang, and Dale Woodford.
\newblock Spanner: Google's globally-distributed database.
\newblock In {\em OSDI}, 2012.

\bibitem{CremonaEtAl:17:Hybrid}
Fabio Cremona, Marten Lohstroh, David Broman, Edward~A. Lee, Michael Masin, and
  Stavros Tripakis.
\newblock Hybrid co-simulation: it's about time.
\newblock {\em Software and Systems Modeling}, 18:1655--1679, November 2017.

\bibitem{dahmann1997department}
Judith~S Dahmann, Richard~M Fujimoto, and Richard~M Weatherly.
\newblock The department of defense high level architecture.
\newblock In {\em Proceedings of the 29th conference on Winter simulation},
  pages 142--149, 1997.

\bibitem{GilbertLynch:02:CAP}
Seth Gilbert and Nancy Lynch.
\newblock Brewer's conjecture and the feasibility of consistent, available,
  partition-tolerant web services.
\newblock {\em ACM SIGACT News}, page 33(2), June 2002.

\bibitem{Kirsch:12:LET}
Christoph~M Kirsch and Ana Sokolova.
\newblock The logical execution time paradigm.
\newblock In {\em Advances in Real-Time Systems}, pages 103--120. Springer,
  2012.

\bibitem{kleppmann2015critique}
Martin Kleppmann.
\newblock A critique of the {CAP} theorem, 2015.
\newblock arXiv:1509.05393 [cs.DC].

\bibitem{Lamport:84:TimeStamps}
Leslie Lamport.
\newblock Using time instead of timeout for fault-tolerant distributed systems.
\newblock {\em ACM Transactions on Programming Languages and Systems},
  6(2):254--280, 1984.

\bibitem{Lamport:78:Time}
Leslie Lamport, Robert Shostak, and Marshall Pease.
\newblock Time, clocks, and the ordering of events in a distributed system.
\newblock {\em Communications of the ACM}, 21(7):558--565, 1978.

\bibitem{Lee:21:Determinism}
Edward~A. Lee.
\newblock Determinism.
\newblock {\em ACM Transactions on Embedded Computing Systems (TECS)},
  20(5):1--34, July 2021.

\bibitem{LeeEtAl:7:DiscreteEvents}
Edward~A. Lee, Jie Liu, Lukito Muliadi, and Haiyang Zheng.
\newblock Discrete-event models.
\newblock In Claudius Ptolemaeus, editor, {\em System Design, Modeling, and
  Simulation using {Ptolemy II}}. Ptolemy.org, 2014.

\bibitem{LeeSan:98:TaggedSignal}
Edward~A. Lee and Alberto Sangiovanni-Vincentelli.
\newblock A framework for comparing models of computation.
\newblock {\em IEEE Transactions on Computer-Aided Design of Circuits and
  Systems}, 17(12):1217--1229, 1998.

\bibitem{Lee:20:Coevolution}
Edward~Ashford Lee.
\newblock {\em The Coevolution: The Entwined Futures of Humans and Machines}.
\newblock MIT Press, Cambridge, MA, 2020.

\bibitem{Liao:97:Scenic}
Stan Liao, Steve Tjiang, and Rajesh Gupta.
\newblock An efficient implementation of reactivity for modeling hardware in
  the {Scenic} design environment.
\newblock In {\em Design Automation Conference}. ACM, Inc., 1997.

\bibitem{LoBelloEtAl:19:TSN}
Lucia Lo~Bello and Wilfried Steiner.
\newblock A perspective on {IEEE} time-sensitive networking for industrial
  communication and automation systems.
\newblock {\em Proceedings of the {IEEE}}, 107(6):1094--1120, 2019.

\bibitem{Lohstroh:EECS-2020-235}
Marten Lohstroh.
\newblock {\em Reactors: A Deterministic Model of Concurrent Computation for
  Reactive Systems}.
\newblock PhD thesis, EECS Department, University of California, Berkeley, Dec
  2020.

\bibitem{Lohstroh:2019:CyPhy}
Marten Lohstroh, {\'I}{\~n}igo {\'I}ncer~Romeo, Andr\'es Goens, Patricia
  Derler, Jeronimo Castrillon, Edward~A. Lee, and Alberto
  Sangiovanni-Vincentelli.
\newblock Reactors: A deterministic model for composable reactive systems.
\newblock In {\em 8th International Workshop on Model-Based Design of Cyber
  Physical Systems (CyPhy'19)}, volume LNCS 11971. Springer-Verlag, 2019.
\newblock in press.

\bibitem{LohstrohEtAl:21:Towards}
Marten Lohstroh, Christian Menard, Soroush Bateni, and Edward~A. Lee.
\newblock Toward a lingua franca for deterministic concurrent systems.
\newblock {\em ACM Transactions on Embedded Computing Systems (TECS)},
  20(4):Article 36, May 2021.

\bibitem{Lynch:96:IOAutomata}
N.~A. Lynch.
\newblock {\em Distributed Algorithms}.
\newblock Morgan Kaufmann, 1996.

\bibitem{Maler:92:Hybrid}
Oded Maler, Zohar Manna, and Amir Pnueli.
\newblock From timed to hybrid systems.
\newblock In {\em Real-Time: Theory and Practice, REX Workshop}, pages
  447--484. Springer-Verlag, 1992.

\bibitem{Mattern:88:virtualtime}
Friedemann Mattern.
\newblock Virtual time and global states of distributed systems.
\newblock In Michel Cosnard, Patrice Quinton, Michel Raynal, and Yves Robert,
  editors, {\em Parallel and Distributed Algorithms}, pages 215--226.
  North-Holland, 1988.

\bibitem{Chandy:86:DDE}
Jayadev Misra.
\newblock Distributed discrete event simulation.
\newblock {\em ACM Computing Surveys}, 18(1):39--65, 1986.

\bibitem{RAO199449}
B.S.Y. Rao and P.~Varaiya.
\newblock Roadside intelligence for flow control in an intelligent vehicle and
  highway system.
\newblock {\em Transportation Research Part C: Emerging Technologies},
  2(1):49--72, 1994.

\bibitem{SchneiderSvH13}
Christian Schneider, Miro Sp{\"o}nemann, and Reinhard von Hanxleden.
\newblock Just model! -- {P}utting automatic synthesis of node-link-diagrams
  into practice.
\newblock In {\em Proceedings of the IEEE Symposium on Visual Languages and
  Human-Centric Computing (VL/HCC '13)}, pages 75--82, San Jose, CA, USA,
  September 2013.

\bibitem{SchwarzMattern:94:CausalConsistency}
Reinhard Schwarz and Friedemann Mattern.
\newblock Detecting causal relationships in distributed computations: in search
  of the holy grail.
\newblock {\em Distributed Computing}, 7:149--174, 1994.

\bibitem{Sepasgozar:22:NetworkTraffic}
Sanaz~Shaker Sepasgozar and Samuel Pierre.
\newblock Network traffic prediction model considering road traffic parameters
  using artificial intelligence methods in vanet.
\newblock {\em IEEE Access}, 10:8227--8242, 2022.

\bibitem{YuVahdat:06:CAP}
Haifeng Yu and Amin Vahdat.
\newblock The costs and limits of availability for replicated services.
\newblock {\em ACM Transactions on Computer Systems}, 2(24):70--113, February
  2006.

\bibitem{Zeigler:1997:DEVS}
Bernard~P Zeigler, Yoonkeon Moon, Doohwan Kim, and George Ball.
\newblock The devs environment for high-performance modeling and simulation.
\newblock {\em IEEE Computational Science and Engineering}, 4(3):61--71, 1997.

\bibitem{ZhangEtAl:18:AWStream}
Ben Zhang, Xin Jin, Sylvia Ratnasamy, John Wawrzynek, and Edward~A. Lee.
\newblock {AWStream}: Adaptive wide-area streaming analytics.
\newblock In {\em Proceedings of SIGCOMM}, Budapest, Hungary, August 20-25
  2018. ACM.

\bibitem{Zhao:07:PTIDES}
Yang Zhao, Edward~A. Lee, and Jie Liu.
\newblock A programming model for time-synchronized distributed real-time
  systems.
\newblock In {\em Real-Time and Embedded Technology and Applications Symposium
  (RTAS)}, pages 259 -- 268. IEEE, 2007.

\end{thebibliography}

\end{document}